\documentclass[twocolumn]{aastex631}

\newcommand{\kms}{$\rm km~s^{-1}$}
\newcommand{\cmc}{$\rm cm^{-3}$}

\newcommand{\vshock}{$v_{shock}$}
\newcommand{\map}{\texttt{MAPPINGS}}
\newcommand{\mapv}{\texttt{MAPPINGS V}}
\newcommand{\jr}{\texttt{CR code}}
\newcommand{\sii}{{[S\sc{ii}]}}
\newcommand{\nii}{{[N\sc{ii}]}}
\newcommand{\oi}{{[O\sc{i}]}}
\newcommand{\oiii}{{[O\sc{iii}]}}
\newcommand{\ariii}{{[Ar\sc{iii}]}}
\newcommand{\kmss}{$\rm km~s^{-1}$ }
\usepackage{natbib}
\usepackage{multirow}
\usepackage{booktabs}
\usepackage{amsmath}
\usepackage{CJK}

\begin{document}
\begin{CJK*}{UTF8}{gbsn}

\title{{Dialogue Concerning the Two Shock Codes}}

\author[0000-0003-0401-3688]{Yifei Jin (金刈非)}
\email{yfjsci@gmail.com}
\affiliation{Westlake University, 
600 Dunyu Road, Xihu District, 
Hangzhou, Zhejiang 310030 PR China}
\affiliation{Center for Astrophysics, Harvard \& Smithsonian, 
60 Garden Street, 
Cambridge, MA 02138 USA}

\author[0000-0002-7868-1622]{John Raymond}
\affiliation{Center for Astrophysics, Harvard \& Smithsonian, 
60 Garden Street, 
Cambridge, MA 02138 USA}
\email{jraymond@cfa.harvard.edu}

\begin{abstract}

In this paper, we summarize the shock physics and the treatment of radiative transfer in two well-established shock codes -- the \map\ code \citep{Dopita1976,Binette1985,Sutherland1993} and the Cox/Raymond code (hereafter \jr) \citep{Cox1972,Raymond1976,Raymond1979}.
We compare the ionization states, temperatures, electron densities, and the energy transportation of the shock models with shock velocities of 50, 110, 150 and 300 \kms .
In summary, both codes adopt the Rankine-Hugoniot flow equation to describe the shock flows, giving the same shock physical properties at the immediate area behind shock fronts.
The different treatments of radiative transfer in these two codes leads to somewhat different computation of the ionization and thermal structures of shocks, as well as the emission-line fluxes.
This work highlights the importance of the delicate treatment of photoionization in shock models, providing insight of the future development of shock codes, such as the 3D shock codes. 

\end{abstract}


\section{Introduction} \label{sec:intro}

Detailed models of the emission spectra of radiative shock waves date back to the work of \cite{Cox1972}.  As more powerful computers became available, models by \cite{Dopita1976}, \cite{Shull1979} and \cite{Raymond1979} added more elements, more atomic rates and more detailed radiative transfer. Over the years, the code of \cite{Dopita1976} evolved into the MAPPINGS code, and the \cite{Raymond1979} code was modified to deal with high densities and other special cases.  A model for shocks in supernova (SN) ejecta was developed by \cite{itoh81}, and subsequent authors added more capabilities \citep{borkowski90, sutherland95, blair00,koo13}.

As with all models, there are some approximations.  Most of the models assume that the thermal conduction term in the energy equation can be ignored, because the transverse B field will be compressed in the shock so that the component of the temperature gradient along the B field will be small.  If that is not the case, thermal conduction can be significant \citep{Borkowski89}.  Most models assume that the elemental abundances are constant throughout the postshock cooling region, but the gradual liberation of refractory elements such as C, Si, Ca and Fe as grains are sputtered affects both the flow structure and the emission spectrum \citep{Slavin25}.  Most tabulated models assume equal electron and ion temperatures, though most models are capable of dealing with unequal electron and ion temperatures with energy transfer between the species by Coulomb collisions, as in \cite{Borkowski89} and models of shocks in SN ejecta \citep{itoh81, blair00}.  The models we discuss here assume a 1D, steady flow, but thermal instabilities affect shocks faster than about 150 \kmss \citep{Binette1985, Innes92, Sutherland03}.  The models of radiative shocks also ignore the pressure of cosmic rays and their contribution to the ionization rate.  This is probably a good assumption for radiative shocks because the particle acceleration efficiency of the relatively slow shocks that are typically radiative is small.  However, reacceleration and compression of energetic particles in the postshock flow might make this component important.  

The applications of shock models are broad.  The theoretical emission-line spectra created from the shock codes provide a reference to interpret the observational spectra by identifying shocks \citep{Yan2018, Flury2024,O'Sullivan2025}, calibrating the chemical abundance \citep{Dopita2016, Dopita2019,D'Eugenio2025}, and measuring the shock speeds and densities \citep{Long2018,Caldwell2025}. 
The theoretical shock emission-line spectra libraries are created by using different parameters for many models.
\cite{Dopita1995} run the original MAPPINGS code to explore the spectral signatures of the low-density fast radiative shocks.
This grid is later developed by \cite{Allen2008} and is widely applied to  Herbig-Haro objects (HH objects), supernova remnants, AGN jets and outflows, and galactic merger systems, because of its large coverage of the range of shock velocities, preshock densities, chemical abundance and magnetic field strengths. \cite{Hartigan87} presented a grid of models from the \jr\ code that has been widely used for interpreting HH object and SNR spectra.  The models can also be used to interpret absorption line spectra \citep{Raymond91}.

Unfortunately, validation of shock models is difficult despite the broad application of shock models.
Comparison of different shock codes is one approach to validation.
The philosophy is to use different codes to run a series of models (benchmarks) with identical set-ups to test different aspects of the numerical codes.
\cite{Ferland1995} gathered the most well-known photoionization and shock codes developers to undertake this benchmark testing 30 years ago, and concluded that the scatter of predicted shock emission-lines is large among the shock codes because of the range of the collisional rates and radiative transfer implemented in the shock codes.

The advent of the James Webb Telescope and the large spatially-resolved database of the Local Volume Mapper project in the fifth Sloan Digital Sky Survey also open the new window to undertake innovative shock physics by providing a huge amount of emission-line spectra.
Also, the shock codes have evolved in the past 30 years by improving the calculation of the preionization in preshock gas and updating the atomic data.
It is therefore time to create a new comparison of shock codes to align the fundament physics in different codes.
This is also the preparation for the next-generation shock models dealing with 3D time-dependent dynamic flows instead of the 1D steady flow.

In this paper, we run a set of benchmarks by using the \mapv\ \citep{Sutherland2017} and the \jr\ \citep{Raymond2020a}.
In Sections~\ref{sec:codes} to \ref{sec:rt}, we summarize the flow physics, atomic physics and the radiative transfer in the two codes. 
We describe the benchmark models in Section~\ref{sec:benchmark} and the comparison of the products from the two codes from Section~\ref{sec:shockstructure} to \ref{sec:shockspectra}.
We briefly summarize the comparison of the models with the observational data in Section~\ref{sec:compareobs} and the conclusion is in Section~\ref{sec:conclude}

\section{Shock Models} \label{sec:codes}

\subsection{Shock Flow Dynamics} \label{sec:model}

In both the \map\ and \jr codes, the shock flow is described by the Rankine-Hugoniot flow equations and the entropy-producing solution is selected. 
Following \cite{Cox1972}, both the steady shock flow and the shock jump are solved in the same manner by deriving the solution to the compression factor, $x\equiv\rho_1/\rho_0$, between any two points in the flow.
The entire shocked flow can be consistently computed by using a single shock flow solver.

The flow maintains the conversation of mass flux:
\begin{equation}
    \rho_0 v_0 = \rho_1 v_1,
\end{equation}
where $\rho$ is the gas density and $v$ is the gas velocity. 
The subscript numbers refer to any two points, $f_0$ and $f_1$, in the flow.
The compression factor, $x$, is defined as
\begin{equation}
    x \equiv \rho_1/\rho_0 = v_0/v_1.
\end{equation}
The frozen-in magnetic field component perpendicular to the flow velocity is constrained by:
\begin{equation}
    B_0/\rho_0 = B_1/\rho_1.
\end{equation}
There are three pressures, gas pressure $P_{gas}$, ram pressure $P_{ram}$, and magnetic pressure $P_B$ in shocks:
\begin{equation}
    P_{gas} = nkT = \frac{\rho kT}{\mu m} \equiv (\gamma -1)E,
\end{equation}
\begin{equation}
    P_{ram} = \rho v^{2},
\end{equation}
\begin{equation}
    P_B = B^2 / 8\pi,
\end{equation}
where $k$ is the Boltzmann constant, $T$ is the temperature, $n$ is the total gas number density, $\mu$ is the molecular weight per gas particle, $m$ is the atomic mass unit, and $E$ is the internal specific energy.
The conservation of momentum flux is expressed in terms of the pressures:
\begin{equation}
    P_{gas,0} + P_{ram,0} + P_{B,0} = P_{gas,1} + P_{ram,1} + P_{B,1}.
\end{equation}
Finally, the energy conservation is expressed as:
\begin{equation}
\begin{split}
\frac{1}{\rho_0}\left(gP_{gas,0} + \frac{1}{2}P_{ram,0} + 2P_{B,0} - \langle L \rangle \Delta t \right) \\
= \frac{1}{\rho_1}\left(gP_{gas,1} + \frac{1}{2}P_{ram,1} + 2P_{B,1} \right),    
\end{split}\label{eq:flow}
\end{equation}
where $g=\gamma/(\gamma-1)$, and $\langle L \rangle \Delta t$ is the energy loss between $f_0$ and $f_1$ through the energy radiated from shocks.
In \jr , the ionization state of plasma is explicitly tracked as an additional energy term of the mean ionization energy per particle \citep{Cox1972}, while \map\ couples the cooling with the ionization states of plasma without tracking the ionization energy separately.

The competition among the gas pressure $P_{gas}$, shock ram pressure $P_{ram}$ and magnetic pressure $P_B$ is the key to determine whether the flow is shocked. As summarized in \cite{Sutherland2017}, shocks only exist when the flow is supersonic according to the following criteria.
\begin{equation}
     \gamma P_{gas}/P_{ram}<1,
\end{equation}
\begin{equation}
     2 P_{B}/P_{ram}<1.
\end{equation}

In the \jr, the ionization energy is kept as a separate term in the energy equation.  Therefore, changes in the ionization energy are not included in the cooling rate, while they are included in \map .  This difference in bookkeeping only affects the cooling rates that are printed out by the codes.  The energetics and temperatures are not affected.
Neither the \jr\ nor \map\ includes thermal conduction or the energy and pressure terms corresponding to cosmic rays. 
 
\subsection{The Precursor Calculation}

{\it MAPPINGS:}
For shocks with velocity $v_{shock}>40 \rm ~km~s^{-1}$ \citep{Dopita2017}, the EUV photons formed in the cooling zone create a photoionization precursor upstream. 
The photoionization in the precursor changes from fully ionized gas to partially ionized gas
 as the shock velocity declines from fast shocks ($v_{shock}>$100~km~s$^{-1}$) to slow shocks ($v_{shock}\sim$50~km~s$^{-1}$).
In the \map\ code before the forth version \citep[such as \texttt{MAPPINGS III} used in][]{Allen2008}, precursors are assumed to be in equilibrium with constant density, and they are computed separately from the shocked flow. 
Since \texttt{MAPPINGS V} \citep{Sutherland2017}, the full structure of precursor is computed consistently at all velocities by solving the time-dependent ionization and temperature evolution over the entire preshock region by taking a multizone scheme instead of a single constant density approximation.

{\it CR code:}  In the CR code, the radiation field coming out through the shock front is used to compute the ionization state at the input parameter values of $n_0$ and $T_0$ at the shock. 
The ionization state and ionizing radiation field can be used for input for another shock run, and the result iterated.  The resulting equilibrum is not entirely self-consistent, in that the flux of ionizing photons is partly balanced by  the flow of low ionization state material into the precursor.  On the other hand, the assumption that the precursor ionization state is in equilibrium with the radiation field produced by the shock is seldom met.  For instance, when the shock first begins to cool effectively and produce strong EUV radiation, the precursor will be underionized.  Later, as the shock slows down, the recombination time in the preshock gas, around $3\times10^5$ years for a density of 1 $\rm cm^{-3}$, is likely to exceed the dynamical time scale of the SNR, and the precursor will be overionized compared to the ionizing flux.  Therefore, the CR code treats the ionization fractions of H and He as free parameters, though they are often set to the equilibrium values  to minimize the number of free parameters.  \cite{Cox1985} gives a simple method to approximate a shock in partially neutral gas as the sum of emission from a thin zone of cooling and ionization just behind the shock plus the emission from a slower shock in fully ionized gas.

\section{Atomic Data}\label{sec:atom}

{\it MAPPINGS:} The atomic data in \map\ has been synchronized with an updated database since its second version \citep{Sutherland1993}.
\texttt{MAPPINGS V} is based on the CHIANTI 8 database \citep{DelZanna2015} and includes 80,000 cooling and recombination lines, and temperature-dependent spline-based interpolation of collisional transitions.
The charge exchange and low temperature dielectronic recombination are also included as described in \citep{Sutherland1993}.

{\it CR code:} Most of the atomic data were assembled in the 1970's and 1980's, with some updates in more recent years.  Many important rates have not changed much in the intervening time, for instance collisional excitation rates of strong permitted lines and excitation rates and Einstein A values for the strong forbidden lines.  The main systematic difference is that the \jr\ uses ionization rates from Summers \citep{Summers1974}.  They agree with the rates used by CHIANTI and by \map\ at temperatures below the temperature at which the abundance of an ion peaks, and they overestimate the ionization rates at temperatures significantly above that.  Another difference, which only becomes important at density on the order of stellar coronal densities, is that the \jr\ includes the density dependence of the dielectronic recombination rates based on fits to \cite{Summers1974}.

The code includes charge transfer between H and heavier ions, along with low temperature dielectronic recombination.  Collisional excitation by protons is included for many fine structure transitions.  Detailed treatment of [Fe II] emission was added by \cite{Hartigan04} and \cite{Koo16}.

\section{Radiative Transfer}\label{sec:rt}

The treatment of radiative transfer, especially the scattering and absorption of hydrogen resonance lines, is different in the two codes.

{\it MAPPINGS:} In \map, the cooling spectrum is tracked from 10$^{-6}$ eV to 10$^5$ eV with the fixed logarithmic bin size.
Both continuum and line radiation fields are considered and the absorption is computed by the cross-section of each ion.
The transfer of hydrogen and helium resonance lines is assumed to be the linear combination of Case A and Case B based on the local optical depth of Ly$\gamma$, using the method of \cite{Brocklehurst1971} and \cite{Johnson1972}.  
This is because the high temperature and large velocity gradients in shock flows make the hydrogen lines neither Case A or Case B.
The scattering of the other resonance lines is included by increasing their effective path length.
The resonance line absorption coefficient across each slab is increased by a factor of $1/P_{esc}$, where $P_{esc}$ is the escape probability formulism given by \citep{Capriotti1965}.
The effects of thermal broadening and the velocity gradient within each slab are included.

{\it CR code:}  The \jr\ bins the emission spectrum into energy bins, typically on the order of 1 eV, and it computes the absorption by the continuum photoabsorption of the atoms and ions.  The exception is atomic hydrogen, for which it assumes Case B.  That is, all  the higher Lyman lines are absorbed locally, and they scatter enough that they are converted to Ly$\alpha$, the two-photon continuum, and lines of the Balmer and higher series.  In addition, the Lyman continuum is treated in the on-the-spot approximation.  That is, photons emitted by recombination to the ground state are absorbed locally, so in effect only recombination to n=2 and above counts toward the recombination rate.  Both these approximations are very good in the photoionization/recombination zone below 10,000 K.  They are less accurate in the narrow region behind the shock where neutral H is excited and ionized, because the optical depth for both lines and continuum is lower there.

An escape probability formalism is available in the CR code for use in high density shocks, where the escape probability effectively reduces the Einstein A values of permitted lines and allows collisional quenching at densities below those that would apply to optically thin gas.  This option was used for shocks in O star winds \citep{krolik85}.

\section{benchmark models}\label{sec:benchmark}

The shock structure is sensitive to the shock velocity. 
In slow shocks, the ionization state of the preshock gas effectively changes the postshock structures by increasing the ionization states of H, He and He$^{+}$, and altering the cooling efficiency in the postshock region.
In fast shocks, the atoms are ionized to high ionization states, increasing the strength and hardness of the ionization fields radiated from the postshock cooling zones.
The role of this radiation field becomes significant as the ionizing photons radiated upstream and downstream form the precursors and the recombination tails, where many of the optical emission lines are created.
The intermediate speed shocks are in the transition regime, where the subtle differences between the pre-ionized preshock and the ionized postshock strongly change the shock structures and emission line fluxes.

We choose four shock models with velocities of 50 \kms , 110 \kms , 150 \kms , and 300 \kms , covering the slow, intermediate speed and fast shocks.
We do not include the shocks with \vshock$>$300 \kms\ because the thermal instability of these ultra-fast shocks breaks the steady shock flow assumption, making the 1D shock calculation invalid.

The magnetic field strength, gas density and chemical abundance are the other pivotal parameters that influence shock models.
The typical magnetic field strength of ISM is around 10 $\mu$G in the Milky Way \citep{Han2017} and is 1-10 $\mu$G in the nebulae \citep{Harvey-Smith2011}.
We set a magnetic field of 5 $\mu$G as the representative magnetic field strength of the ISM.
The gas density of the models is set to be 5 \cmc\ to represent the broad range of gas densities in the Milky Way \citep{Kennicutt2012}.
We only consider the most abundant 12 coolants following the \cite{Asplund2009} solar abundance set.

\section{Shock Structures}\label{sec:shockstructure}

The ionization and temperature of the preshock gas are determined from the photoionization equilibrium and ionization balance calculation using the upstream ionization field.
The extended precursor region is established if the flux of ionizing photons passing upstream through the shock front exceeds the flux of neutrals approaching the shock.

Once the gas passes through the shock front, the ion and electron temperatures are discontinuously enhanced.
The postshock gas cools down strongly through the collisional cooling due to the high collisional rates.
The population of each ionization state is calculated from the ionization and recombination rate of each ion. 
In these models, the electron-ion temperatures are locked together and a Maxwellian electron energy distribution is assumed.

\begin{figure*}
    \centering
    \includegraphics[width=0.8\linewidth]{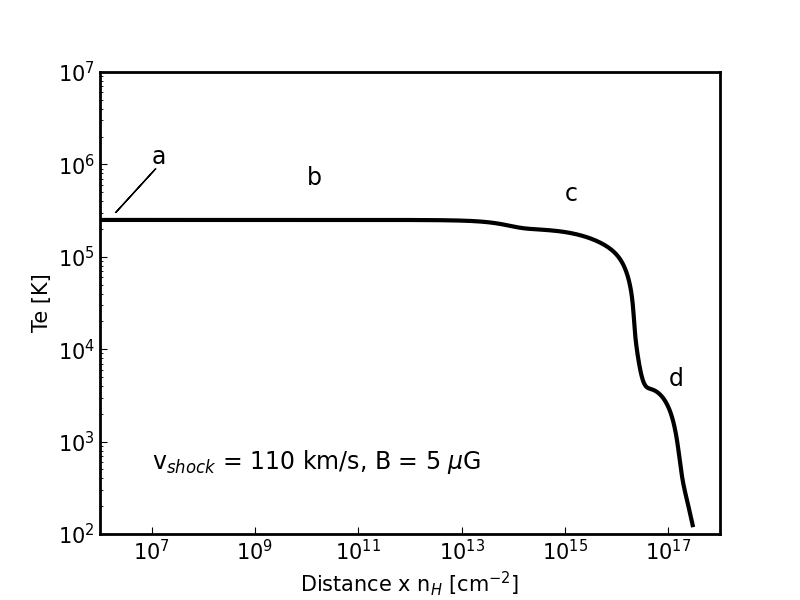}
    \caption{The electron temperature of a shock model as a function of the product of the distance and gas density. The shock velocity is 110~\kms , the preshock gas density is 5~\cmc and the magnetic field is 5~$\mu$G. Four sub-structural zones: a) {\it the immediate postshock region}, b) {\it the relaxing zone}, c) {\it the cooling zone} and d) {\it the recombination zone} are labeled.}
    \label{fig:ShockStructure}
\end{figure*}

The shock structure is identified as four major regions/zones as shown in Figure~\ref{fig:ShockStructure}.
These regions are also described in \cite{Cox1972,Cox1985,Dopita1996} and \cite{Sutherland2017}.

a) \textit{The Immediate Shock Region} -- This is the region next to the shock front. The gas ionization states are those of the preionized preshocked gas but the temperature discontinuously increases on the plasma scale given by the proton Larmor radius or the ion skin depth. 

b) \textit{The Relaxing Zone} -- Here the ionization states of gas approach collisional ionization balance and the gas temperature remains hot, though significant cooling can occur is if the preshock gas is partly neutral \citep{Cox1985}. 
The plasma is optically thin in this zone.  If the shock is much younger than the radiative cooling time, this is the only zone present, and it would be a nonradiative shock or Balmer line filament \citep{Chevalier78}.

c) \textit{The Cooling Zone} -- The temperature drops off from 10$^{5-6}$~K to 10$^4$~K in this zone as the majority of the cooling and emission of the internal ionizing radiation field occurs.
The upstream radiation field preionizes the precursor, and the downstream radiation field further ionizes the cooler gas, forming the photoinization/recombination zone.
In the cooling zone, the ionic populations recombine from their high-ionization states to the intermediate- and low-ionization species.
For instance, the \oiii\ emissions arise when the $\rm O^{+3}$ population recombines to $\rm O^{+2}$.

d) \textit{The Recombination Zone} -- As the plasma recombines and cools down, the downstream EUV photons re-ionize the gas, approaching photoionization-recombination equilibrium.
This region behaves similar to an H{\sc ii} region ionized by a hard ionization field, producing the majority of the optical emission-lines.  
For the shock flow with magnetic fields, the recombination tail is supported by the magnetic field pressure, which determines its extent, density, ionization state and temperature.

\subsection{The Shock Front}

We first compare the solution of the shock front of the two codes by investigating the physical properties before and after the jump.
In the preshock region, we compare the ionization state of H and He because it indicates the strength of the upstream ionization field, which is a reflection of the internally-generated ionizing photons of shocks. 
The H and He ionic populations are also the critical initial condition which determines the cooling in postshock regions.
In the immediate postshock region, we compare the electron temperature, density and the hydrogen density.
These parameters describe the gas properties given by the Rankine-Hugoniot jump conditions, indicating the accuracy of the equation solver implemented in each code.
These solutions are not affected by any processes related to atomic physics, like cooling or ionization, because the energy term in equation~\ref{eq:flow}, $\Delta L = 0$, across the shock front, therefore, the properties of the immediate postshock regions are purely dominated by the flow dynamics. 

The ionization states of H and He of the preshock gas are shown in Table~\ref{tab:ionpreshock}.
In the 50~\kms\ shock flow, the preshock hydrogen and helium are almost neutral with only around 1 percent ionization. 
This is because very few ionizing EUV photons are produced in the postshock region.
The residual EUV arriving at the shock front is not sufficient to ionize the preshock hydrogen and helium.
As the shock velocity increases, the postshock collisional ionization increases, and more ionizing photons are created. 
Therefore, in \mapv\ and \jr, the ionization state of hydrogen increases at the same pace from the neutral state in the 50~\kms\ shock to the full ionization in the 150 \kms\ and 300 \kms\ shocks through the half ionization state in the 110 \kms\ shock.
The helium ionization in these two codes follows the same trend that fraction of ionized helium increases with increasing shock velocity. 
However, we notice that, in the fast shocks (\vshock =150 \kms\ and 300 \kms), the He-ionization field in \mapv\ is softer than that in \jr\ because the He$^{+2}$/He$^{+1}$ ratio in \mapv\ is smaller than the ratio given by \jr .

Table~\ref{tab:ppposthock} shows the physical conditions of the immediate postshock gas.
Both \mapv\ and \jr\ give similar electron temperatures; the 50~\kms\ shock temperature is around 7.5$\times$10$^{4}$~K, the 110~\kms\ shock temperature is 2.5$\times$10$^{5}$~K, the 150~\kms\ shock temperature is 3.2$\times$10$^{5}$~K, and the 300~\kms\ shock has the hottest temperature of 1.2$\times$10$^{6}$~K.
The hydrogen gas densities given by the two codes are consistent to within 1.6\% accuracy.
The electron densities given by the two codes are also consistent within a uncertainty of 7 percent apart from the 50~\kms\ shock.
The 13.6 percent of difference in $n_e$ in the slow shock may caused by the different treatments of radiative transfer of the upstream ionizing radiation fields.

\subsection{The Relaxing Zone}

In the relaxing zone, the gas approaches collisional equilibrium and the temperature is as hot as the immediate postshock region due to the lack of efficient cooling.
In the models presented here, the electrons and the ions share the same temperature, as inferred from observations for shocks in this velocity range \citep{Ghavamian2013, Raymond2023}. 
In this region, the hydrogen is highly ionized so the gas is optical thin for Lyman lines.
The gas in this region is close to the ``Case A'' assumption.
At the transition region from the relaxing zone to the cooling zones, the hydrogen ionization state and the electron density are further elevated with the increasing collisional ionization due to the rising density of the shock flow.

\subsection{The Cooling Zone}

The shock temperature starts to drop after the shocked gas reaches ionization balance and the collisional cooling becomes efficient.
The cooling spectrum radiated from this region going upstream and downstream is the major contributor to the internal ionization field of shock flows.
Therefore, the calculation of cooling processes determines the structures and emission-line spectra of the precursor and the recombination zone.

\begin{figure*}
    \centering
    \includegraphics[width=0.5\linewidth]{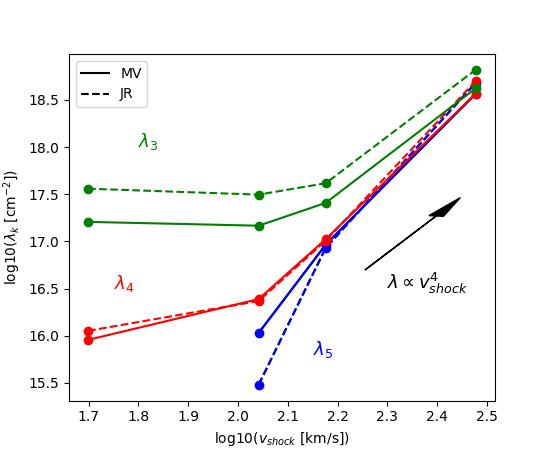}
    \caption{The cooling columns as a function of shock velocities. The cooling column, $\lambda_k$=$\Delta r_k \times n_H$, where $\Delta r_k$ is the cooling length at which the temperature drops to a specific value. The k=3,4,5./ corresponding to $T_e = 10^{3}, 10^{4}$, and $10^{5}$~K. The solid line is the prediction of \mapv\ and the dashed line is the prediction of \jr.}
    \label{fig:coollength}
\end{figure*}

Following \cite{Sutherland2017}, we define the cooling column, $\lambda_k$=$\Delta r_k \times n_H$, to account for the speed of cooling in the shock gas.
The $\Delta r_k$ is the distance from the shock front to the point at which the shock temperature drops to a specific temperature $T_e$, where k=3,4,5 corresponds to $T_e = 10^{3}, 10^{4}$, and $10^{5}$~K.
Figure~\ref{fig:coollength} shows the changes of $\lambda_k$ as a function of shock velocity.
As the shock velocity increases, the temperature drops quickly from 10$^5$~K to 10$^3$~K.
In the fast shock regime of \vshock$>$110~\kms , the collisional ionization dominates the postshock structure until the very end of the shock flow.
Therefore, the cooling length principally reveals the 
cooling time multiplied by 1/4 the shock velocity, where the cooling time is the thermal energy divided by $n_e \times \Lambda(T)$.
The flatness of $\lambda_k$ in Figure~\ref{fig:coollength} at low velocity is because there are very few free electrons at 50 \kms.  
At high velocities, the slope is closer to \vshock$^4$, because there is a \vshock$^2$ from the thermal energy and a \vshock\ from the shock speed and roughly \vshock\ from the lower $\Lambda_{cool}$ at $T_e$ above 100,000~K.

\begin{figure*}
    \centering
    \includegraphics[width=\linewidth]{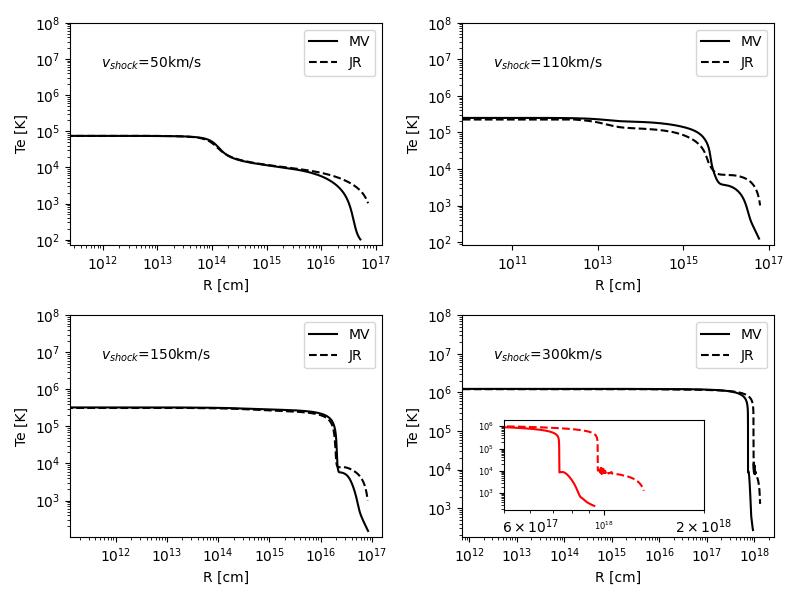}
    \caption{The comparison of the electron temperatures predicted by \mapv\ (solid) and \jr\ (dashed) as a function of the distance to the shock front. In the \vshock=300~\kms\ panel, the small window shows the zoom-in cooling and recombination region between $6\times10^{17}<r<2\times10^{18}$~cm.}
    \label{fig:teprofile}
\end{figure*}

We notice that the \mapv\ shock models cool down faster than the \jr\ models as shown in Figure~\ref{fig:teprofile}.
For the 50~\kms\ shock model, the \jr\ model and the \mapv\ model both start cooling down at 10$^{14}$~cm from 8$\times$10$^4$~K to 10$^{4}$~K but the \jr\ model maintains its temperature at 10$^4$~K longer than the \mapv\ model does.
For the 110~\kms\ and 150~\kms\ shock models, the \jr\ model cools down more efficiently than the \mapv\ model at 10$^{13}$-10$^{14}$~cm, so that the temperature of the \jr\ shock model is cooler than the \mapv\ model between 10$^{13}$~cm and 10$^{16}$~cm.
For the 300~\kms\ shock model, the \mapv\ model starts to cool down at 7$\times$10$^{17}$~cm, which is earlier than the \jr\ model does at $10^{18}$~cm.
Therefore, the 300~\kms\ \jr\ shock has larger cooling zone than the \mapv\ shock.

\subsection{The Recombination Zone}

The recombination zone is the end of shock flows where the typical temperature is around $10^4$~K.
This zone is heated by the EUV ionizing photons produced in the cooling zone.
The magnetic pressure supports the gas and lowers the density, so the thickness of the recombination zone is larger in the faster shocks with stronger magnetic fields.
The gas becomes neutral in the recombination zone, contributing the majority of the emission-lines in shock spectra.

Both \mapv\ and \jr\ create the recombination zones in 110~\kms , 150~\kms\ and 300~\kms\ shocks.
However, the \jr\ models have the hotter and thicker recombination zone than the \mapv\ models, especially in the 300~\kms\ shocks.
As shown in Figure~\ref{fig:coollength}, the $\lambda_3$ of the \jr\ models are systematically larger than the \mapv\ by 0.5~dex over the entire shock velocity range. 
This is because that different atomic data bases give different ionization and recombination rates, leading to the difference in the populations of the high and low ionization states as well as the follow-up cooling speed.

\section{Cooling spectra}\label{sec:cooling}

\begin{figure*}
    \centering
    \includegraphics[width=0.8\linewidth]{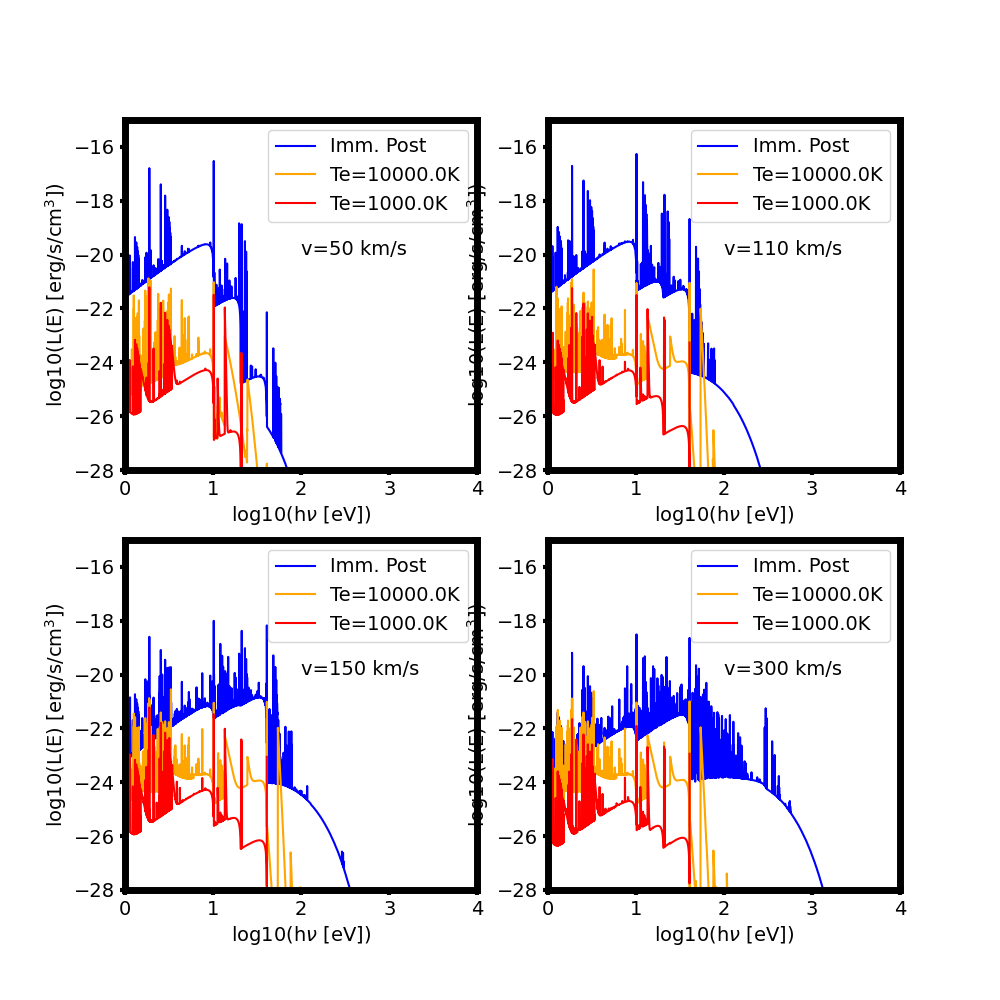}
    \caption{The cooling spectra at the immediate postshock region and the positions of $T_e=10000, 1000$~K. }
    \label{fig:coolsp}
\end{figure*}

The cooling spectra providing the ionization radiation in postshock regions are composed of the free-free continuum, resonance lines and the two-photon continua produced by the metastable levels of H$^0$, He$^0$ and He$^+$.
Figure~\ref{fig:coolsp} shows the cooling spectra at the immediate postshock region, the position of $\rm T_e$=1e4~K and the position of $\rm T_e$=1e3~K. 
As the temperature cools down, the ionizing spectrum softens. 
As with other shock properties, the ionizing spectrum is a function of shock velocity such that the ionizing radiation field is harder in faster shock flows. 
The 300~\kms\ shock has more ionizing emission-lines than the slower shock models.

\section{Shock spectra}\label{sec:shockspectra}

\begin{figure*}
    \centering
    \includegraphics[width=0.8\linewidth]{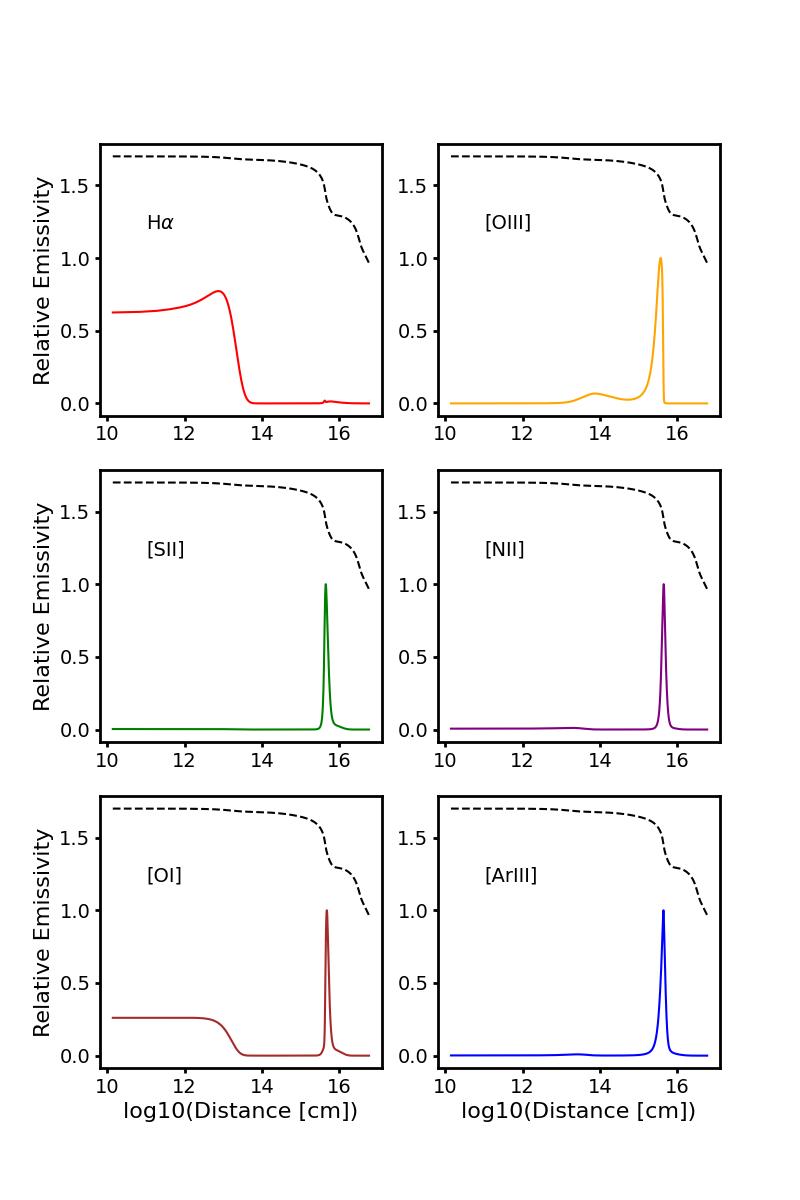}
    \caption{The distribution of emission-line emissivities in the postshock regions of the 110~\kms\ \mapv\ shock model. The solid lines are the relative emissivities of each emission-lines where the peak is to be 1. The dashed line is the re-scaled temperature profile indicating different postshock zones. }
    \label{fig:emiprof}
\end{figure*}

The shock emission-line spectra produced by \mapv\ and \jr\ are the crucial theoretical reference for observers to obtain the properties of the ISM and SNRs.
As shown in Figure~\ref{fig:emiprof}, we investigate the distribution of the emissivities of six diagnostic emission-lines, H$\alpha$, \oiii, \nii, \sii, \oi\ and \ariii. 
Most of the optical emission-lines are radiated from the recombination zone and the end of the cooling zone of the shock flow.
Therefore, the calculation of photoionization in the recombination zone affects these optical emission-line fluxes.

Part of the \oi\ and \oiii\ are radiated before the cooling zone starts.
The \oi\ emissivity close to the shock front is caused by the residual neutral oxygen, O$^{0}$, after the shock front.
The \oiii\ emissivity in the relaxing zone rises after the recombination of O$^{+3}$ to O$^{+2}$ in conjunction with the  increasing density.

In Table~\ref{tab:linflux}, we compare the integrated shock spectra given by \mapv\ and \jr.
The H$\beta$ fluxes are similar given by both codes, which indicates that both \mapv\ and \jr\ produce the similar total ionization energy.

The \mapv\ models give smaller fluxes than the \jr\ models in the emission-lines which are generated in the recombination zone, like the \sii$\lambda\lambda$6717,32, \nii$\lambda$6584, \nii$\lambda$5677, \oi$\lambda$6300 and \ariii$\lambda7751$.
The difference in fluxes of these emission-lines recalls the fact that the \mapv\ and \jr\ codes have different treatments of photoionization, giving different structures in the recombination zone.
The recombination tails in the \mapv\ models are shorter than the \jr\ models, leading to the smaller fluxes of the emission-lines generated in this region.

The \oiii$\lambda$5007 emission predicted by these two codes have different dependences on the shock velocity. 
The \oiii\ emission turns on from 110~\kms\ shock model.
In the \jr\ models, the \oiii/H$\beta$ ratio remains relatively flat in the 150~\kms\ and 300~\kms\ shock models.
In the \mapv\ models, the \oiii/H$\beta$ ratio declines by a factor of 2 from the 110~\kms\ shock model to the 300~\kms\ shock model.

\subsection{The Dependence on Magnetic Field}

\begin{figure*}
    \centering
    \includegraphics[width=0.8\linewidth]{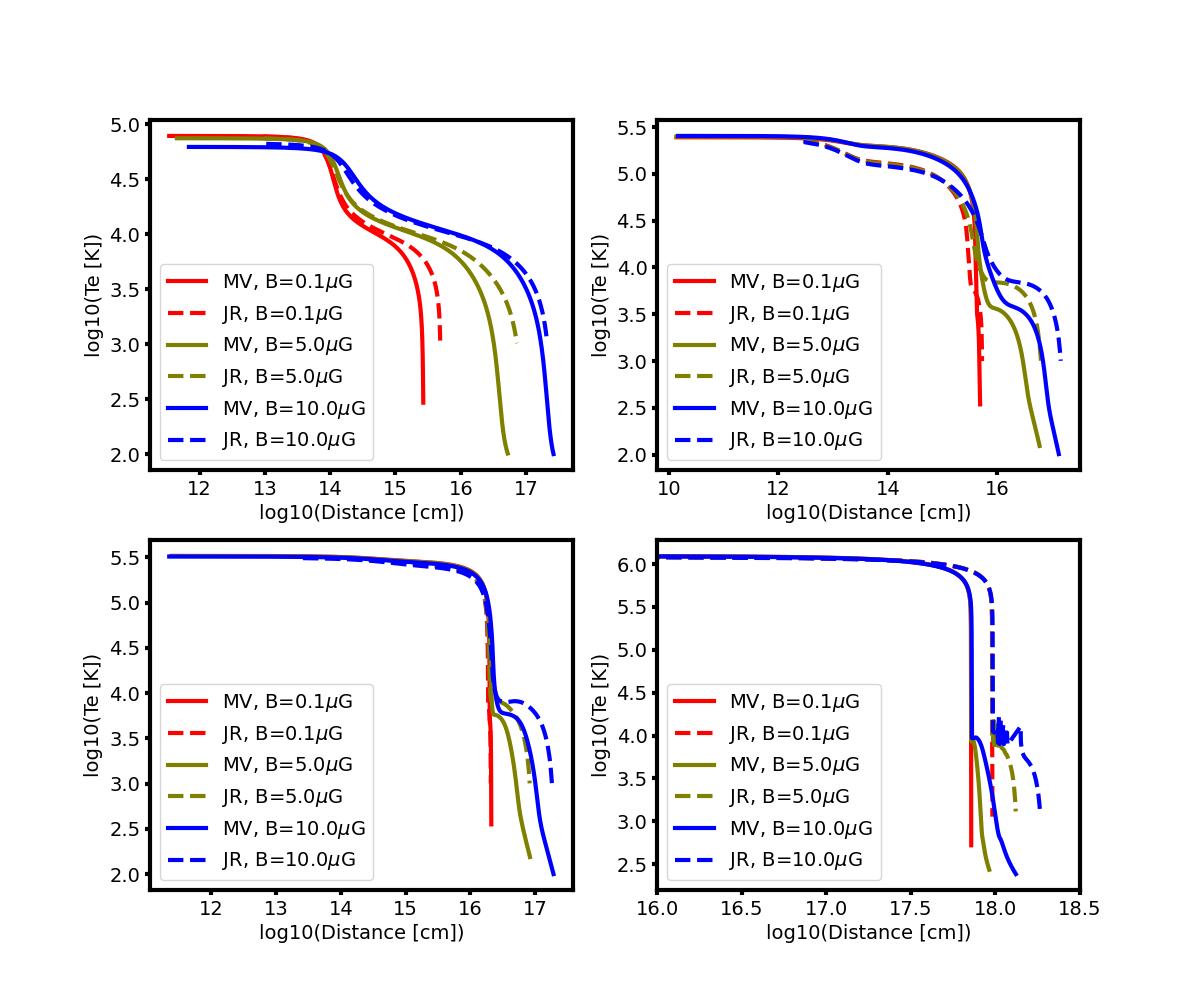}
    \caption{The temperature profiles of the \jr\ shock models (dashed lines) and the \mapv\ shock models (solid lines) with 0.1~$\mu$G, 5~$\mu$G and 10~$\mu$G magnetic strength.}
    \label{fig:magprof}
\end{figure*}

We create two sets of models with the magnetic strength of 0.1~$\mu$G and 10~$\mu$G as the low- and high-magnetic strength comparison models.
As shown in Figure~\ref{fig:magprof}, the stronger transverse magnetic pressure reduces the density, and therefore collisional rates as well as the cooling rate, extending the size of the recombination zone.
However, the \jr\ models have a stronger response to the magnetic field than the \mapv\ models. 
The recombination regions extend farther in the \jr\ models than those in the \mapv\ models as the magnetic strength increases.
This is probably because the \map\ code has more accurate treatment of the cooling and ionization, such as the refined frequency sampling of the ionizing spectra, than the \jr .
%Therefore, the energy restored in magnetic fields is efficiently radiated through coolings, hedging the impact of magnetic field. 
%because of stronger photoionization. 

\begin{figure*}
    \centering
    \includegraphics[width=\linewidth]{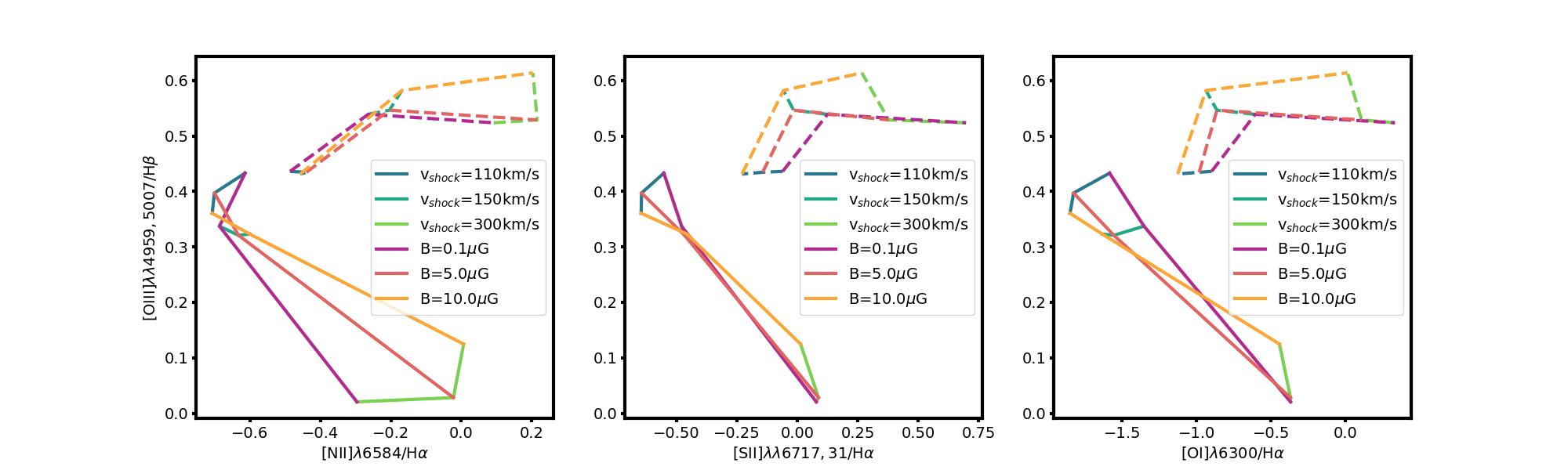}
    \caption{BPT diagram of the 110, 150 and 300\kms\ shock models.}
    \label{fig:bpt}
\end{figure*}

In Figure~\ref{fig:bpt}, we compare the ratios of \oiii/H$\beta$, \nii/H$\alpha$, \sii/H$\alpha$ and \oi/H$\alpha$, which are the so-called BPT diagnostic lines \citep[Baldwin, Phillips, \& Terlevich diagnostic lines][]{Baldwin1981}.
The \oiii/H$\beta$ ratios are mainly determined by shock velocities and the influence of the magnetic strength is only a secondary factor.
For the other BPT-diagnostic line ratios, the \jr\ models and the \mapv\ models give the same dependence on the magnetic strength, where the line ratios decrease as the magnetic strength increases.
For the 300~\kms\ models, the difference of the effect of magnetic field is negligible when the magnetic strength is smaller than 5~$\mu$G.

\section{Comparison with Observations}\label{sec:compareobs}

Ideally, one could compare these models with observations to determine their accuracy.  In practice, there are generally more free parameters than observables.  The biggest problem is that there are very few observed radiative shocks whose shock speeds are known independently.  An exception is the western region of the Cygnus Loop, where proper motions have been measured from HST observations by \cite{Raymond2020a}.  The distance to the Cygnus Loop was believed to be 735$\pm$25 pc \citep{fesen18}, but recent work has shown that it could be as far as 800 pc \citep{ritchey24}.  even so, the shock speeds derived from the proper motions and distances are good to about 15\%.  The shock speeds range from about 90 \kms~to about 160 \kms.  \cite{Raymond2020a} found that the average spectra agreed reasonably well with the \jr\ models, but with significant discrepancies for the low ionization states.

A more complete comparison was carried out by \cite{Slavin25}.  That paper combined UV spectra from HST/COS with fluxes in optical emission lines obtained by integrating the fluxes in HST narrow band images over the COS apertures.  They examined 6 positions whose shock speeds span the 90 to 160 \kms ~range.  The goal of that work was to investigate dust destruction in shocks, so it combined the \jr\ model with dust destruction calculations from \cite{jones96} for a self-consistent computation of the ionization states, cooling and emission line spectra.  The models matched the UV spectra quite well, but they predicted far too much emission in  the lowest temperature lines, [O I] and [S II].  This is almost certainly because the COS aperture size is similar to the cooling length for these shocks, and the positions were chosen to be bright in [O ~III].  Therefore, most of the photoionization/recombination zone was missed.  The HST observations in general show that the morphology of the [O~III] emission is much different than that of the H$\alpha$ or [S~II] lines, largely because of the development of turbulence behind the shocks.  This calls into question the applicability of 1D steady-flow models to any astrophysical shocks.  However, when an observation takes in a large area, it seems that the average spectrum can be meaningfully compared with the shock models simply because both the models and the actual shocks basically just conserve matter, momentum and energy. However, it may be necessary to average the spectra over a range of shock speeds \citep{vancura92}.

\section{Conclusion}\label{sec:conclude}

We compare the shock models predicted by two well-established shock codes, the \jr\ and \mapv .
The shock flows in both codes are described by the Rankine-Hugoniot flow equation with the solutions proposed by \cite{Cox1972}.
The flow equation solvers behave identically in both codes, giving the similar solutions of shock jumps.
The treatments of the radiative transfer and the photoionization are different in the two codes, leading to some uncertainty of the shock emission-line spectra.

We create a set of four shock models with the shock velocity of 50~\kms, 110~\kms, 150~\kms\ and 300~\kms\ to cover the cases from slow shocks to fast shocks through intermediate-speed shocks.
These models can be used as the benchmarks in future developments of shock codes.

Both codes produce the entire shock structures, which are the immediate postshock region, the relaxing zone, the cooling zone and the recombination tail.
The shock structures produced by \jr\ are slightly different from the \mapv\ because of the more approximate treatments of the radiative transfer and the photoionization.
\mapv\ tracks the ionizing radiation field with around 7000 bins, producing accurate ionic population of up-to 30 elements.
This refined treatment of photoionization gives more accurate predictions of the temperature of the recombination tails as well as the shock emission-lines.
In these benchmark models, we see that the \jr\ models cool down faster than the \mapv\ models in the relaxing zone and the cooling zone.
On the other hand, the \jr\ models have hotter and more extended recombination tails than the \mapv\ models.

The significance of photoionization in shock models is amplified in predicted shock emission-line spectra.
The \oiii/H$\beta$ ratios given by the two codes show distinguishable dependence on shock velocities.
The \oiii/H$\beta$ ratio is subtly dependent on shock velocity in the \jr\ models but the \mapv\ models show that the \oiii/H$\beta$ ratio decreases more strongly with shock velocity in the faster shocks.
The pronounced difference in line ratios given by the two codes highlights the intrinsic uncertainty of the interpretation of shock emission-line spectra, particularly for the extragalactic studies.

The fluctuations of density and magnetic field strength caused by shock-turbulence interaction is ignored in both codes. 
More realistic density structures will affect the radiation field and the balance between the gas turbulent motion and  magnetic pressure.
Our current work already demonstrates the importance of the accurate calculation of radiative transfer and photoionization in predicting shock emission-line spectra.
We strongly call for the development of the next-generation three-dimensional shock code.

\begin{acknowledgments}

YFJ acknowledge the support from Prof. Yong Shi at Westlake University.

\end{acknowledgments}

\begin{table*}[]
    \caption{Chemical Abundance adopted from \cite{Asplund2009}}
    \centering
    \begin{tabular}{c|c}
\hline
\hline    
Element & log(X/H) \\
\midrule
 H &   0.00 \\
  He & -1.07 \\
  C & -3.57 \\
  N & -4.17 \\
  O & -3.31 \\
 Ne & -4.07 \\
 Mg & -4.40 \\
 Si & -4.49 \\
 S & -4.88 \\ 
 Ar & -5.60 \\
 Ca & -5.66 \\
 Fe & -4.50 \\
 \hline
 \hline
    \end{tabular}
    \label{tab:my_label}
\end{table*}

\begin{table*}
\caption {Preshock Ionization States} \label{tab:ionpreshock} 
\begin{center}
\begin{tabular}{c|cc|cc|cc|cc}
\hline
\hline
 \multicolumn{1}{c}{}   &  \multicolumn{2}{c}{50 \kms} & \multicolumn{2}{c}{110 \kms}  & \multicolumn{2}{c}{150 \kms} & \multicolumn{2}{c}{300 \kms} \\
\midrule
 \multicolumn{1}{c}{}   &  JR & MV & JR & MV & JR & MV & JR & MV \\
\midrule
H I  &     0.988  & 0.989  & 0.557  & 0.573  & 0.010  & 0.010  & 0.001  & 0.002 \\
He I  &     1.000  & 0.990  & 0.580  & 0.577  & 0.000  & 0.102  & 0.000  & 0.000 \\
He II  &    0.000  & 0.010  & 0.420  & 0.418  & 0.670  & 0.854  & 0.000  & 0.114 \\
\hline
\hline
\end{tabular}
\end{center}
\end{table*}

\begin{table*}
\caption {Physical Properties of the Immediate Postshock Region} \label{tab:ppposthock} 
\begin{center}
\begin{tabular}{c|ccc|ccc|ccc|ccc}
\hline
\hline
 \multicolumn{1}{c}{}   &  \multicolumn{3}{c}{50 \kms} & \multicolumn{3}{c}{110 \kms}  & \multicolumn{3}{c}{150 \kms} & \multicolumn{3}{c}{300 \kms} \\
\midrule
 \multicolumn{1}{c}{}   &  JR & MV & $\Delta$ & JR & MV & $\Delta$ & JR & MV & $\Delta$ & JR & MV & $\Delta$ \\
\midrule
$T_e$  &    7.53e4 & 7.48e4 & 0.7$\%$  &   2.43e5  & 2.51e5 & 3.2$\%$ &   3.22e5  & 3.27e5 & 1.5$\%$ &  1.23e6 & 1.24e6 & 0.8$\%$ \\
$n_H$  &    17.2   & 17.4   & 1.2$\%$  &   19.1    & 18.8   & 1.6$\%$ &    19.3   & 19.5   & 1.0$\%$ &   19.9  & 19.8   & 0.5$\%$ \\
$n_e$  &    0.19   & 0.22   & 13.6$\%$ &   8.45    & 8.76   & 3.5$\%$ &    19.70  & 20.90  & 5.7$\%$ &   21.40  & 20.00 & 7.0$\%$  \\
\hline
\hline
\end{tabular}
\end{center}
\end{table*}

\begin{figure*}
    \centering
    \includegraphics[width=\linewidth]{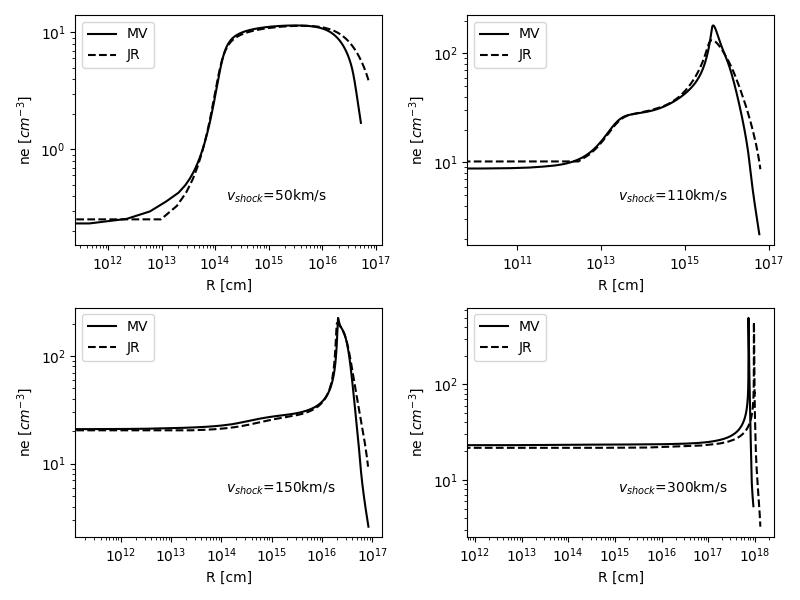}
    \caption{The comparison of the electron density predicted by \mapv\ (solid) and \jr\ (dashed) as a function of the distance to the shock front.}
    \label{fig:neprofile}
\end{figure*}

% 50 km/s numbers updated 05/27

\begin{table*}
\caption {Fluxes of Important Emission Lines} \label{tab:linflux} 
\centering
\begin{tabular}{cc|cc|cc|cc|cc}
\hline
\hline
\multicolumn{2}{c}{Lines} &  \multicolumn{2}{c}{50 \kms} & \multicolumn{2}{c}{110 \kms}  & \multicolumn{2}{c}{150 \kms} & \multicolumn{2}{c}{300 \kms} \\
\midrule
 H$\beta=$100    & ($\rm \AA$)    &  JR & MV & JR & MV & JR & MV & JR & MV \\
\midrule
O IV    &  1400.00  &  0.00  &  0.00  & 35.00  & 141.37   & 319.00 &257.43 &  157.00 &  109.97 \\
C IV  &   1549.00  &  0.00  &  0.00 &  276.00 & 1258.63 &  580.00 &  473.40 &  281.00 &  222.41 \\
He II    &  1640.00  &  0.00  &  0.00  & 27.00  & 65.53  & 72.00  & 40.20 &  266.00 &  140.41 \\
O III D  &  1664.00  &  0.00  &  0.00 &  110.00 &  149.52 &  172.00 &  137.07 &   71.00  & 55.36 \\
Si III  & 1890.00  &  0.00  &  0.05 &  144.00  & 41.74  & 81.00  & 28.91  & 44.00  & 12.44 \\
C III D  & 1907.00  &  1.00  &  0.30 &  439.00 &  525.16 &  231.00 &  203.24 &  152.00 &  109.94 \\
O II  &   3727.00  & 43.00  & 26.25 &  470.00 &  525.18 &  634.00 &  388.21 & 1171.00 &  688.14 \\
Ne III D  & 3867.00  &  0.00  &  0.00  & 37.00  & 48.91  & 42.00  & 38.71 &  109.00  & 77.84 \\ 
O III  & 4363.00  &  0.00  &  0.00  & 18.00  & 23.50  & 25.00  & 21.07  & 12.00  &  8.82 \\
He II    &  4686.00  &  0.00  &  0.00  &  0.00  &  8.04  &  0.00  &  5.35  &  0.00  & 21.69 \\
O III D & 5007.00  &  0.00  &  0.00 &  272.00 &  336.26 &  352.00 &  282.09 &  294.00 &  144.07 \\
N II D    & 5677.00  &  0.00  &  0.00  &  3.00  &  0.02  &  3.00  &  0.01  &  7.00  &  0.00 \\
He I  &   5876.00  &  0.00  &  0.31  & 13.00  & 11.90  & 13.00  &  8.90  & 17.00  & 17.64 \\
O I D     &  6300.00  & 54.00  & 44.08  & 39.00  &  6.56  & 57.00  & 12.01 &  276.00 &  171.92 \\
H$\alpha$   & 6562.00 &  355.00 &  408.66 &  281.00 &  319.74 &  312.00 &  309.79 &  295.00 &  296.23 \\
N II  &   6584.00  & 30.00  & 14.16 &  102.00  & 86.46 &  195.00  & 99.67 &  391.00 &  383.43 \\
S II   & 6717.00  & 77.00  & 47.02 &  117.00  & 41.28 &  171.00  & 60.11 &  270.00 &  197.63 \\
S II   & 6732.00  & 53.00  & 32.78  & 85.00  & 33.16 &  130.00  & 49.95 &  207.00 &  172.14 \\
Ar III   & 7751.00  &  0.00  &  0.00  &  5.00  &  1.13  &  6.00  &  0.84  & 26.00  &  4.08 \\
S III D   & 9530.00  &  4.00  &  2.08  & 37.00  & 31.59  & 60.00  & 24.42 &  240.00 &  152.00 \\
Ar II    & 6.98 $\rm\mu m$  &  3.00  &  1.32  &  3.00  &  5.71  &  6.00  &  7.19  &  6.00  &  7.99 \\
Ne II    &   12.0 $\rm\mu m$  &  31.00  &  0.88  & 43.00  & 48.32  & 56.00  & 54.29  & 55.00  & 54.13 \\
Ne III   &   15.5 $\rm\mu m$  &  0.00  &  0.00  & 12.00  & 15.71  & 24.00  & 17.81 &  124.00 &  156.62 \\
O IV  & 25.9 $\rm\mu m$  &  0.00  &  0.00  &  2.00  & 11.89  & 11.00  & 17.59  &  8.00  &  8.90 \\
Si II    &   34.8 $\rm\mu m$ &  473.00 &  390.59 &  446.00 &  500.74 &  490.00 &  435.43 &  724.00 &  687.07 \\ 
\midrule
 \multicolumn{2}{c|}{H$\beta$ (erg cm$^{-2}$ s$^{-1}$)}     &    4.08e-06 &    4.74e-06  &   4.00e-05  &   3.98e-05  &   10.9e-05  &   9.00e-05  &   2.99e-04 &    3.80e-04 \\
\hline
\hline
\end{tabular}
\end{table*}

\begin{table*}
\caption {Ly$\alpha$/Ly$\beta$ versus H$\alpha$/H$\beta$} \label{tab:caseab} 
\begin{center}
\begin{tabular}{c|cc|cc|cc|cc}
\hline
\hline
 \multicolumn{1}{c}{}   &  \multicolumn{2}{c}{50 \kms} & \multicolumn{2}{c}{110 \kms}  & \multicolumn{2}{c}{150 \kms} & \multicolumn{2}{c}{300 \kms} \\
\midrule
 \multicolumn{1}{c}{(H$\beta$=100)}   &  JR & MV & JR & MV & JR & MV & JR & MV \\
\midrule
Ly$\beta$  &    0    & 33.9    &  0 & 324.6   &   0 & 23.3    & 0   & 11.1   \\
Ly$\alpha$  &  8237. & 9076.3  & 4941.  & 4679.1  &  3773.  & 2917.2  &  4750.  & 2972.6 \\
H$\alpha$  &  355.     & 408.7    & 281.  & 319.7   & 312.   & 310.8   & 295.   & 296.2  \\
\midrule
Ly$\alpha$/Ly$\beta$  &   $\infty$    & 267.7  &  $\infty$   & 14.4  &  $\infty$   & 125.2  & $\infty$   & 267.8 \\
H$\alpha$/H$\beta$  &  3.5       & 4.1    & 2.8    & 3.2   & 3.1    & 3.1    & 3.0   & 3.0 \\
\hline
\hline
\end{tabular}
\end{center}
\end{table*}

\clearpage

\bibliography{biblio}{}
\bibliographystyle{aasjournal}

\end{CJK*}

\end{document}